\documentclass[iop,revtex4]{emulateapj}
\usepackage{graphics,graphicx,natbib}
\usepackage{ textcomp }

\newcommand{\ee}[1]{\mbox{${} \times 10^{#1}$}}
\newcommand{\eten}[1]{\mbox{$10^{#1}$}}
\newcommand{\as}{\mbox{\arcsec}}

\newcommand{\water}{H$_2$O}

\newcommand{\hh}{\mbox{{\rm H}$_2$}}
\newcommand{\nn}{\mbox{{\rm N}$_2$}}
\newcommand{\ammonia}{\mbox{{\rm NH}$_3$}}
\newcommand{\nnhp}{\mbox{\rm N}$_{2}$\rm{H}$^+$}

\begin{document}

\title {The Effects of Initial Abundances on Nitrogen in Protoplanetary Disks}
\author{Kamber R. Schwarz and Edwin A. Bergin}
\affil{Department of Astronomy, University of Michigan, 500 Church Street, Ann Arbor, MI 48109, USA}

\begin{abstract}
The dominant form of nitrogen provided to most solar system bodies is currently unknown, though available measurements show that the detected nitrogen in solar system rocks and ices is depleted with respect to solar abundances and the interstellar medium.
We use a detailed chemical/physical model of the chemical evolution of a protoplanetary disk to explore the evolution and abundance of nitrogen-bearing molecules. Based on this model we analyze how initial chemical abundances, provided as either gas or ice during the early stages of disk formation, influence which species become the dominant nitrogen bearers at later stages. We find that a disk with the majority of its initial nitrogen in either atomic or molecular nitrogen is later dominated by atomic and molecular nitrogen as well as \ammonia\ and HCN ices,
where the dominant species varies with disk radius.
 When nitrogen is initially in gaseous ammonia, it later becomes trapped in ammonia ice except in the outer disk where atomic nitrogen dominates. For a disk with the initial nitrogen in the form of ammonia ice the nitrogen remains trapped in the ice as NH$_{3}$ at later stages. 
The model in which most of the initial nitrogen is placed in atomic N best matches the ammonia abundances observed in comets. 
Furthermore the initial state of nitrogen influences the abundance of \nnhp, which has been detected in protoplanetary disks.
Strong \nnhp\ emission is found to be indicative of an \nn\ abundance greater than $n_{\nn}/n_{\hh}>\eten{-6}$, in addition to tracing the CO snow line. Our models also indicate that NO is potentially detectable, with lower N gas abundances leading to higher NO abundances. 
\end{abstract}

\section{Introduction}
Most solar system bodies are depleted in nitrogen relative to the Sun and the ISM. \citet{Pontoppidan14} compare the CNO abundances in various solar system bodies relative to silicon. Comets Halley and Hale-Bopp, which likely formed in the outer disk ($R > 10$ AU), where there is thought to be less chemical reprocessing, are depleted in nitrogen by less than an order of magnitude with respect to solar values. In contrast the nitrogen content of meteorites is between one and three orders of magnitude below the amount of nitrogen that was available, as traced by the Sun \citep{Asplund09}.
In fact, comets and meteorites exhibit greater depletion in nitrogen than in other volatiles such as oxygen and carbon \citep{Pontoppidan14}. In this context the Earth is extremely depleted in nitrogen with an abundance ratio more than five orders of magnitude below the solar abundance. Most of the known nitrogen resides in the atmosphere in the form of \nn, however, it is possible that much of Earth's nitrogen was locked in its interior \citep{Roskosz13}. 
The $\mathrm{^{15}N/^{14}N}$ ratio of the terrestrial surface, which includes the atmosphere and oceans, as well as the crust, agrees with that of chondrites, suggesting that both the Earth's surface and meteorites obtained their nitrogen from the same reservoir \citep{Marty12}.
It remains unclear what the dominant nitrogen-bearing species was upon delivery to the young Earth, whether it was carried by organics in meteorites, \nn\ from the solar nebular gas, or \ammonia/organics in cometary bodies \citep{Epstein87,Wyckoff91,Owen01,Kawakita07}.

The Cronian satellites Titan and Enceladus both show evidence of rich \nn\ atmospheres \citep{Niemann05, Waite06}. 
Isotopic abundances in Titan's atmosphere were measured by the Huygens probe and
the low $^{36}$Ar/$^{14}$N ratio along with the absence of detectable of $^{38}$Ar, Kr and Xe provides circumstantial evidence that Titan first received its nitrogen in a less volatile form, such as \ammonia\ \citep{Niemann05}. The nitrogen could later be converted to the more volatile \nn, possibly by impacts during the period of late heavy bombardment \citep{Sekine11}.

Which nitrogen-bearing molecules end up in solar system bodies depends on the chemical composition of the protoplanetary disk at the time of planetesimal formation.
In this paper we show that the dominant bearer of nitrogen is highly dependent on initial chemical abundances. 
There continues to be some uncertainty in regards to the nitrogen partitioning in dense molecular pre-stellar cores, which represent the initial conditions. 
This uncertainty stems from the difficulty of detecting many of the most probable nitrogen reservoirs: N, \nn, and nitrogen-bearing ices. N and \nn\ are not directly observable in the dense ISM.  Instead their abundances must be inferred from observations of trace molecules such as N$_{2}$H$^+$, which is a reaction product of \nn.  
\ammonia\ ice abundances are derived from absorption features. However, the measured \ammonia\ ice abundances are uncertain due to the blending of particular absorption features with those of water and silicates \citep{Oberg11}.

Spectroscopic observations of ices toward low-mass young stellar objects as part of the Spitzer ``Cores to Disks'' program find that on average 10\% of the total nitrogen is contained in known ices, primarily \ammonia, NH$_{4}^{+}$, and XCN (OCN$^-$), though in some sources the percentage is as high as 34\% \citep{Oberg11}. 
 The remaining nitrogen is posited to reside in the gas phase as either atomic N or \nn. \citet{Womack92} estimate \nn\ abundances in multiple dense molecular cores using observations of \nnhp. They find an average fractional abundance of 4$\times10^{-6}$ with respect to \hh, or $6\%$ of the total nitrogen reservoir assuming solar abundances \citep{Asplund09}.
\citet{Maret06} find  the gas phase nitrogen to be primarily in atomic as opposed to molecular nitrogen for the prestellar core B68, though based on their models the main nitrogen-bearing species is \ammonia\ ice.
Similarly,  \citet{Daranlot12} conclude that 45\% of the total elemental nitrogen in dense clouds is in the form of \ammonia\ ices on grains, though the abundances they predict are larger than those observed.
Le Gal et al. (2014) modeled the gas phase nitrogen chemistry in dark clouds. Starting with nitrogen all in gas phase N they find that once a steady state is reached the dense core contains equal abundances of N and \nn\ with a small fraction of the initial nitrogen in other species. 

\ammonia\ has long been a tracer of dense cores \citep[e.g.][]{Benson89}. Using detections of \ammonia\ inverse transitions toward five starless cores, \citet{Tafalla02} calculated \ammonia\ gas abundances ranging $4.0\ee{-9}$ to $1.0\ee{-8}$ with respect to \hh.
\ammonia\ gas has also been observed in absorption several thousand AU from the Class 0 protostar IRAS 16293-2422 with 
an abundance relative to \hh\ of  $\approx 3.6\times \eten{-7} - 6.5\times \eten{-7}$
\citep{HilyBlant10}. 
Finally, \citet{LeGal14} find that the steady state abundances of nitrogen hydrides in their chemical model were in good agreement with those observed toward IRAS 16293-2422. 
In sum, in prestellar cores, gas phase \ammonia\ is clearly not a major nitrogen reservoir, however as discussed earlier there is a significant amount of \ammonia\ potentially present in the ices.
\citet{Visser11} modeled the chemical evolution of a collapsing protostar from a pre-stellar core to a disk, finding that the disk begins with most of the nitrogen in gas phase \nn\ with minor contributions from N, \ammonia, and NO.

Despite detections in these earlier stages of star formation there are currently no published detections of \ammonia\ gas
in protoplanetary disks, though upper limits exist for several T Tauri stars in the near-infrared \citep{Salyk11,Mandell12}.
However, both CN and HCN have been detected in multiple protoplanetary disks \citep[e.g.][]{Oberg11b,Guilloteau13} and resolved \nnhp\ emission is seen in the T Tauri system TW Hya \citep{Qi13b} while unresolved emission is seen toward several more systems \citep{Dutrey07,Oberg10,Oberg11b}. 

Thus a variety of potential initial distributions of the total nitrogen reservoir could be provided to the forming disk. It is possible that a substantial fraction of the nitrogen could be in the form of N gas, \nn\ gas, or \ammonia\ ices. Alternatively, if the source is warm enough during collapse, a significant amount of \ammonia\ could be provided to the disk as \ammonia\ gas. In this paper we perform simulations of the disk chemistry in which we make four different initial abundance assumptions: (1) most of the nitrogen is in atomic form; (2) most of the nitrogen arrives in the disk in molecular form; (3) the nitrogen arrives as \ammonia\ ice; (4) most of the nitrogen arrives as \ammonia\ gas.

\section{Model}
We use the disk chemistry model of \citet{Fogel11} to explore the effects of different initial abundances.
We first set the physical structure of the disk, including temperature, density, and dust properties. Next we perform UV and X-ray radiative transfer, assuming all radiation is from the central star. The model then calculates chemical abundances based on a network of 5903 chemical reactions and specified initial abundances. The specifics of the model are discussed in greater detail below.

\subsection{Physical Structure}
\begin{figure}
\centering
    \includegraphics[width=1.0\textwidth]{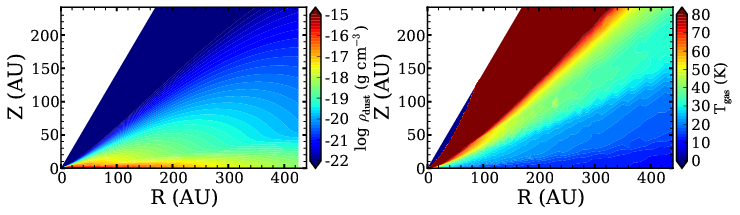}
  \caption{ Disk model dust density and temperature structure.}
  \label{physicalproperties}
\end{figure}

We use the two-dimensional, azimuthally symmetric disk physical structure adopted by \citet{Cleeves13}. This model is designed to represent a `typical' T-Tauri disk based on the transition disk observations presented by \citet{Andrews11}. The density structure is fixed and of the form:
\begin{equation}
\Sigma_{g}(R)=\Sigma_{c}\left(\frac{R}{R_{c}}\right)^{-1} \exp{\left(-\frac{R}{R_{c}}\right)}.
\end{equation}
Here $\Sigma_{c}$ and $R_c$ are the characteristic surface density and radius, taken to be 3.1 g cm$^{-2}$ and 135 AU respectively.
The disk is a settled disk, $\epsilon$=0.1
, where $\epsilon$ is the dust-to-gas mass ratio of dust grains in the upper disk relative to the standard value of 0.01 \citep{DAlessio06}. 
Smaller $\epsilon$ values indicate less dust in the upper disk and more dust in the midplane. Two types of dust are included: small, micron size grains and larger millimeter size grains. The grain distribution is given by: 
\begin{equation}
\rho_{small} = \frac{(1-f)\Sigma}{\sqrt{2\pi}R h} \exp{\left[-\frac{1}{2}\left(\frac{Z}{h}\right)^{2}\right]},
\end{equation}
\begin{equation}
\rho_{large} = \frac{f\Sigma}{\sqrt{2\pi}R\chi h} \exp{\left[-\frac{1}{2}\left(\frac{Z}{\chi h}\right)^{2}\right]},
\end{equation}
and
\begin{equation}
h(r)=h_{c}\left(\frac{R}{R_{c}}\right)^{\psi}.
\end{equation}
The height profile given by Equation (4) is applied to both the small dust grains and the gas, with a characteristic scale height of $h_c$ = 12 AU and $\psi$ = 0.3. The scale height for the large grains is smaller by a factor of $\chi = 0.2$. 85\% of the total dust mass is in the large grains such that $f$ = 0.85. The dust density and temperature distributions are shown in Figure \ref{physicalproperties}.

\subsection{Radiation Field}
The FUV field from the central star, including Ly$\alpha$ radiation, and the stellar X-ray field were generated using a Monte Carlo radiative transfer and scattering model as described by \citet{Bethell11a,Bethell11b}.
We adopt the FUV spectrum to be that measured for TW Hya \citep{Herczeg02,Herczeg04}. We assume a thermal X-ray spectrum between 1 and 10 keV with an integrated X-ray luminosity of \eten{30} erg s$^{-1}$, which is typical for a T Tauri star \citep{Glassgold97}. We then compute the X-ray attenuation using the cross-sections of \citet{Bethell11b}.

Ly$\alpha$ radiation contains $\sim80\%$ of the total FUV flux \citep{Herczeg04,Bergin03}. Additionally several species have photodissocation cross sections close to Ly$\alpha$. For these reasons, Ly$\alpha$ radiation cannot be ignored, as is the case for the weaker UV emission lines \citep{Fogel11}. In addition to scattering off dust grains, Ly$\alpha$ photons will first isotropically scatter off of hydrogen atoms on the top of the disk surface.  This layer scatters a fraction of the Ly$\alpha$ radiation more directly towards the midplane, allowing the Ly$\alpha$ radiation greater penetrating power than the whole of the FUV continuum and lines beyond Ly$\alpha$ \citep{Bethell11a}.

\subsection{Reaction Network}

We use the chemical model of \citet{Fogel11}, which is based on the gas-phase reaction network of the Ohio State University Astrophysical Chemistry Group \citep{Smith04} and modified to include the updated reaction rates of \citet{McElroy13}. This network does not include the expansions for reactions at high temperatures from \citet{Harada10}. Thus the predictions for the inner edge of the disk may change. However, in this paper we focus on the abundances of molecules with high volatility at radii beyond a few AU.

The chemical code is run at 74 radii with each radius broken into 45 vertical zones. The grid spacing is logarithmic in radius and linear in angle. 
The chemistry in each vertical zone is run independently, except for the considerations needed for self-shielding, which is treated vertically, with no mixing between zones. A pseudo two-dimensional result is obtained by running the model for many radii. 
The model includes photodesorption, photodissociation, freeze out, grain surface reactions, gas phase ion and electron reactions, and cosmic ray and stellar X-ray ionization as well as self-shielding of \hh\ and CO.  
Cosmic rays are assumed to strike the disk vertically with an unshielded ionization rate of $1.3\ee{-17}$ s$^{-1}$.
The photodissociation rates depend on the strength of the radiation field at a given point in the disk and the molecule's cross section. 
Grain surface reactions are limited to the formation of \hh, \water, \ammonia, and CH$_{4}$ via successive hydrogenation. The self-shielding and grain surface reactions are time dependent and the chemistry is run for 3 Myr. 

\subsection{Initial Conditions}
We present the results from four chemical models, each with different initial conditions as listed in Table 1. Model N uses the initial abundances of \citet{Fogel11}, with most nitrogen originating in atomic form. None of the initial nitrogen is in ices. 
Most of the oxygen is provided as either CO, $n_{\mathrm{CO}}/n_{\mathrm{H}}=1\ee{-4}$, or water ice, $n_{\mathrm{H_{2}O(gr)}}/n_{\mathrm{H}}=2.5\ee{-4}$, with a small fraction of the oxygen in other gas phase species.
These values are for the model molecular cloud of \citet{Aikawa99}, which is in good agreement with observed molecular abundances in cores \citep[e.g.][]{Terzieva98}.

In Model \ammonia(gr) most of the nitrogen is in \ammonia\ ice and there is initially no atomic nitrogen. This is a simplification of the \citet{Maret06} and \citet{Daranlot12} models which predict large \ammonia\ ice abundances. Model \ammonia\ differs from Model \ammonia(gr) in that it starts with \ammonia\ in the gas phase. This model is only viable if the collapse of the initial molecular cloud liberates \ammonia\ from grains. Finally, Model \nn\ starts with the nitrogen primarily in \nn. As discussed above, gas phase N and \nn\ are not directly observable, making it difficult to observationally determine the relative gas phase abundances. Model \nn, when compared to Model N, allows us to determine whether the partitioning between N and \nn\ has an observable effect on the chemistry at later stages.
We emphasize that the variations in the initial conditions of our models do not represent the most likely nitrogen distribution in the disk. Rather, they are extreme examples meant to illustrate the effect the initial nitrogen distribution has on the chemistry at later stages.

\section{Model Results}
\subsection{Model N}
\subsubsection{Radial and Vertical Structure}
\begin{figure}
\centering
    \includegraphics[width=1.0\textwidth]{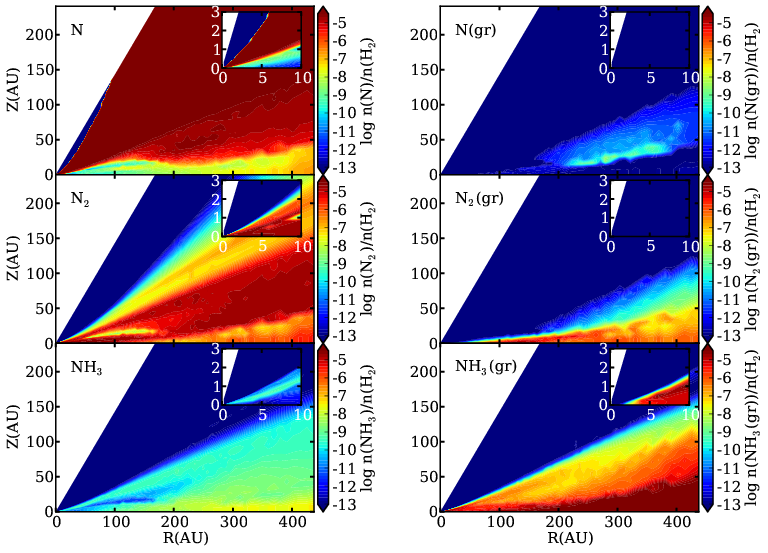}
  \caption{ Abundances for Model N relative to molecular hydrogen (part 1). The inset shows the inner disk.  X(gr) indicates the abundance of species X on dust grains.}
  \label{standardcontours1}
\end{figure}

\begin{figure}
\centering
    \includegraphics[width=1.0\textwidth]{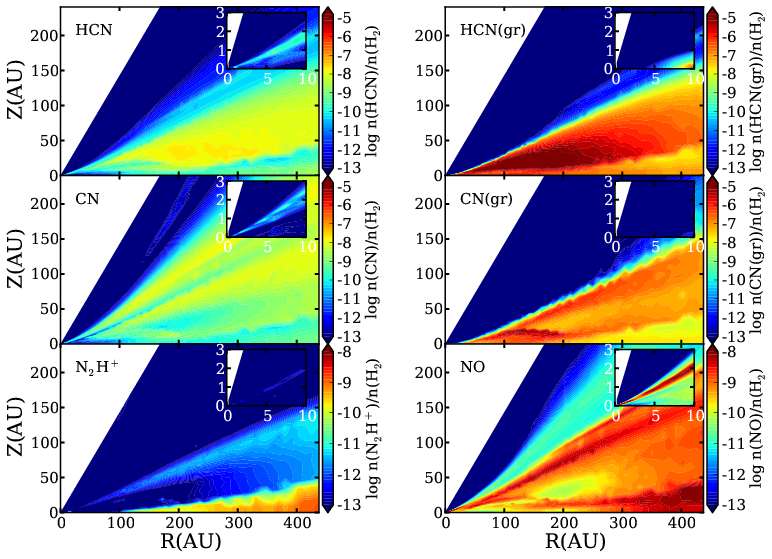}
  \caption{ Abundances for Model N relative to molecular hydrogen (part 2). The inset shows the inner disk.  X(gr) indicates the abundance of species X on dust grains.}
  \label{standardcontours2}
\end{figure}

The following results are for a chemical time of $\sim$1 Myr unless otherwise stated.
The abundances of the major nitrogen species are shown for a cross section of the disk in 
Figures \ref{standardcontours1} and \ref{standardcontours2}. 
In the cold midplane ($Z=0$) the thermal desorption rates are low, allowing many species to freeze out onto ices. 
Above the midplane the ices have evaporated and the chemistry is able to process N into \nn\ as well as \ammonia\ ice. This region corresponds to the `warm molecular layer' \citep{Aikawa02,Bergin07,Henning13,Dutrey14}. At $R$ = 200 AU the warm molecular layer extends vertically from $Z$ $\ge$ 60 AU.  In the upper, photon dominated region of the disk molecules are destroyed quickly and atomic nitrogen dominates. 

\begin{figure}
\centering
    \includegraphics[width=1.0\textwidth]{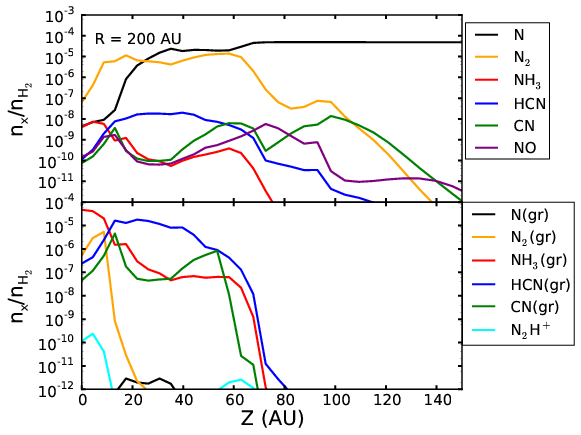}
  \caption{ Abundances of nitrogen-bearing species at a radius of 200 AU after $8.9\times 10^{5}$ yr for Model N.}
  \label{zcut}
\end{figure}

\begin{figure}
\centering
    \includegraphics[width=1.0\textwidth]{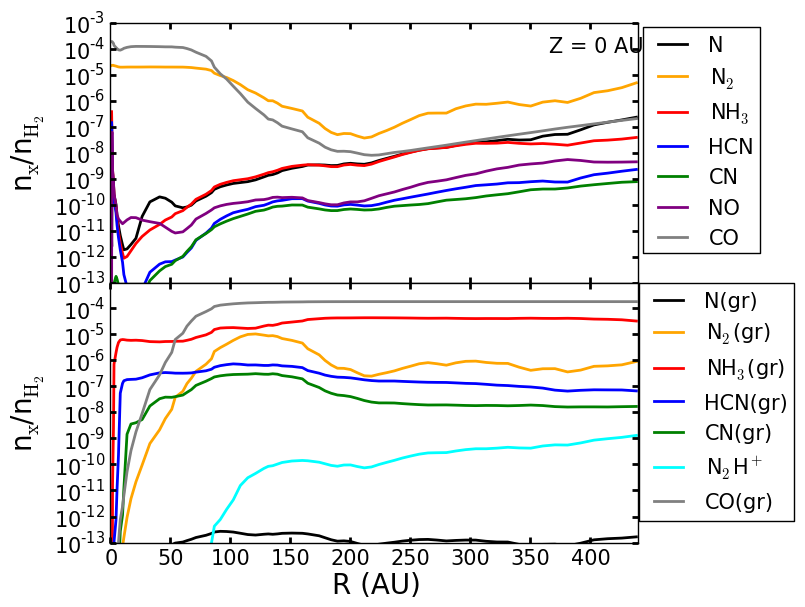}
  \caption{ Radial abundance profiles at the midplane for Model N.}
  \label{snow}
\end{figure}

Figure \ref{zcut} shows the vertical structure of nitrogen carriers at a radius of 200 AU. This far from the central star,
\nn\ ice is only abundant close to the midplane where the gas temperature is low, $T <$ 20 K, and molecules are shielded from radiation from the central star. However, the midplane is dominated by \ammonia\ ice. The majority of the \ammonia\ ice in the midplane is formed on the grains, rather than resulting from freeze out of \ammonia\ gas. Closer to the star \nn\ ice evaporates and HCN ice, which has a higher binding energy, is present in greater abundance (see Figure \ref{snow}). 

Inside $R$ = 200 AU chemical processing converts the N gas into \nn\ in the midplane. At the disk surface N is also able to form \ammonia, CN, and HCN gas. For $R <$ 10 AU these molecules are able to remain in the gas phase in the surface layers, while for $R >$ 10 AU they adsorb onto dust grains. 

The highest concentration of \ammonia\ and \ammonia\ ice in the disk is in the range $Z=\pm$  20 AU except in the outer disk, where \ammonia\ ice remains highly abundant at $Z=\pm$  50 AU (Figure \ref{standardcontours1}). In addition to forming on grains, \ammonia\ is the end result of a series of ion-neutral reactions and forms quickly in these regions. Due to its high binding energy, \ammonia\ ice acts as a sink, leaving little \ammonia\ in the gas phase. 
However, near the midplane there is a residual abundance of \nn\ and \ammonia. Cosmic ray desorption prevents \nn\ from completely adsorbing onto grains. Instead, \nn\ remains active in the gas phase chemistry, allowing the continued formation of \ammonia. 
This chemistry is sensitive to the binding energies of molecules such as \nn\ on grains and to the presence/absence of ionizing photons. However, these binding energies are often uncertain and the presence of cosmic rays in the midplane has been called into question \citep{Cleeves13}. This uncertainty will be discussed in more detail in Section 3.3.

CN and HCN gas phase abundances are low in the cold midplane region (Figure \ref{standardcontours2}). These molecules are formed via gas phase reactions and only exist in abundance in the warm molecular layer. After forming, CN and HCN quickly adsorb onto dust grains. Thus CN and HCN ices also exist in the molecular layer. 
At $R$ = 200 AU, HCN ice becomes the dominant form of nitrogen at heights between 10 and 30 AU (Figure \ref{zcut}). Because much of the CN goes into forming HCN and \nn\ before it can freeze out, a similar peak is not observed in CN ice.

For $R >$ 100 AU, \nnhp\ traces the midplane (Figure \ref{standardcontours2}). Inside 100 AU, \nnhp\ is destroyed via gas phase reactions with CO, which does not exist in abundance in the midplane for $R >$ 100 AU.
NO, a precursor molecule to \nn, is most abundant in the surface layers ($Z\sim50-100$ AU at $R=200$ AU), where the main formation and destruction mechanisms are:
\begin{equation}
\mathrm{N + OH \rightarrow NO + H}
\end{equation}
and
\begin{equation}
\mathrm{N + NO \rightarrow N_2 + O}.
\end{equation}
In the surface layers OH is slightly more abundant than NO, allowing the creation of NO to outpace its destruction.

\subsubsection{Radial Variation in the Midplane}
Figure \ref{snow} illustrates the snow lines for nitrogen-bearing species in the midplane. At a radius of 100 AU in the midplane, \nn\ and N have evaporation temperatures of  14.8 K and 15 K respectively. These temperatures were calculated using Eqn. 5 from \citet{Hollenbach09} and the binding energies in Table 1. In comparison, the evaporation temperatures of HCN and \ammonia\ are 38 K and 57 K. While it is fairly easy to free N and \nn\ from grains in the midplane, HCN and \ammonia\  remain locked in ices outside of 17 AU and 5 AU respectively.
This range of evaporation temperatures results in a series of sublimation fronts. 
The dearth of ices at small radii does not correspond with an increase in the gas phase abundance of the same species. Instead, inside 150 AU the majority of the nitrogen makes its way into \nn\ via chemical processing. This process is discussed in more detail in Section 3.1.3. 

Early in the chemical evolution of the disk, most of the HCN gas in the midplane either freezes out onto grains or is destroyed via gas phase reactions with ions. Because of its high binding energy, it remains on grains for radii greater than 17 AU. The same process occurs for the \ammonia\ and CN gas, resulting in a substantial fraction of the total nitrogen locked in HCN and \ammonia\ ices in the midplane beyond 17 AU.

The resulting reservoir of moderately volatile ices could be used to form more complex molecules, thus changing the picture presented above. \ammonia\ and HCN are both needed to form aminomethanol, a precursor to glycine, in ices via Strecker synthesis \citep{Danger12}. The same ices are also capable of forming hydroxyacetonitrile. In addition hydrogenation of HCN can result in the formation of methylamine
 on grain surfaces \citep{Theule11}.
 
Ices in the presence of UV radiation have been shown to form more complex molecules \citep[e.g.][]{Bernstein02}. For example, irradiation of \ammonia\ ice at 10 K can lead to the formation of NH$_{2}$\ and N$_{2}$H$_{4}$ while \nn\ becomes N$_{3}$ \citep{Gerakines96}. Especially in the warm molecular layer, which is more transparent to UV photons that the midplane, the abundance of \ammonia, \nn, and HCN ices could decreases, corresponding to an increase in more complex species.
 
Turbulent mixing would likely transport ices from the midplane to the warmer upper regions of the disk, where desorption would remove the molecules from grains \citep{Furuya14}. Once in the gas phase molecules such as \ammonia\ would be dissociated, with the nitrogen eventually going into either \nn\ or N depending on the local disk temperature. 
Additionaly, planetesimal drift and advection would bring ice coated grains to smaller radii \citep{Wiedenschilling93,Ciesla07}.
There they would evaporate, likely contributing to the gas phase \nn\ abundance.

\subsubsection{Time Evolution}
\begin{figure}
\centering
    \includegraphics[width=1.0\textwidth]{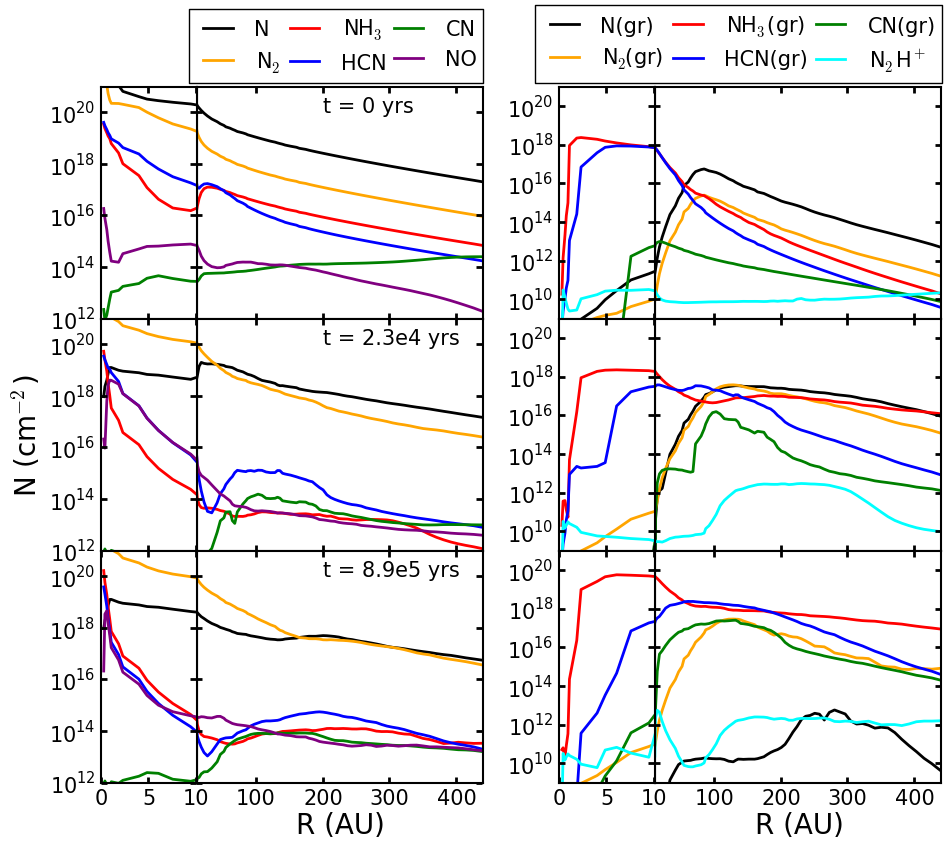}
  \caption{ Column densities for the most abundant nitrogen-bearing species in our Model N. The change in scaling on the x axis at 10 AU is to better show behavior in the inner disk.}
  \label{standardtime}
\end{figure}

To explore time dependent behavior throughout the disk the abundances are collapsed to radial column density plots by integrating over the vertical direction (Figure \ref{standardtime}). These plots provides the most complete information at a given stage in addition to allowing easy examination of the time evolution of the chemistry.

At early stages, atomic N, where the majority of the nitrogen initially resides, and \nn\ dominate.
There is a general evolution of N becoming \nn\ on fairly short time scales via gas phase reactions with CN and NO.
These are both two step reactions:
\begin{equation}
\mathrm{N + CH \rightarrow CN + H}
\end{equation}
\begin{equation}
\mathrm{N + CN \rightarrow N_2 + C}
\end{equation}
and
\begin{equation}
\mathrm{N + OH \rightarrow NO + H}
\end{equation}
\begin{equation}
\mathrm{N + NO \rightarrow N_2 + O},
\end{equation}
though other reactions can also create CN and NO. 
The C and O are released when CO reacts with He$^{+}$ \citep{Bergin14}.
 Beyond 50 AU these reactions primarily take place in the warm molecular layer. Inside 50 AU the same reactions place a majority of the initial N into \nn\ early on.
 At  $t$ = 8.9$\times10^5$ years in our model the N gas column density still surpasses that of \nn\ beyond $R$ = 200 AU. Inside of 200 AU the presence of gas phase CO allows for more O to be present in the midplane. Some of this O goes into OH, leading to the formation of \nn\ as discussed previously.  
The evaporation timescale increases with temperature, so as time passes larger radii begin to be affected by evaporation. For N the evaporation timescale in the midplane at 180 AU is $1.7\times10^5$ years. This, in conjunction with the formation of \ammonia\ on grains, destroys the N snow line at late stages.

Once it freezes out onto grains, N ice is quickly converted to \ammonia\ ice. Over time this depletes the N ice reservoir.
More of the nitrogen becomes locked in \ammonia\ and HCN ices, with HCN ices becoming the dominant nitrogen-bearing species between 60 AU and 170 AU for late stages. 
Closer to the central star HCN is more likely to be photo-dissociated before it can freeze out onto the grains.

As more of the total nitrogen becomes trapped in ices the abundances of the molecules needed to form \ammonia, such as NH$_{4}^{+}$, drop and the formation rate of gas phase \ammonia\ slows. This is also true for the precursors of HCN. Thus there is little change in \ammonia\ ice and HCN ice abundances at late stages.

\subsection{Changes in Initial Conditions}
\begin{figure}
\centering
    \includegraphics[width=1.0\textwidth]{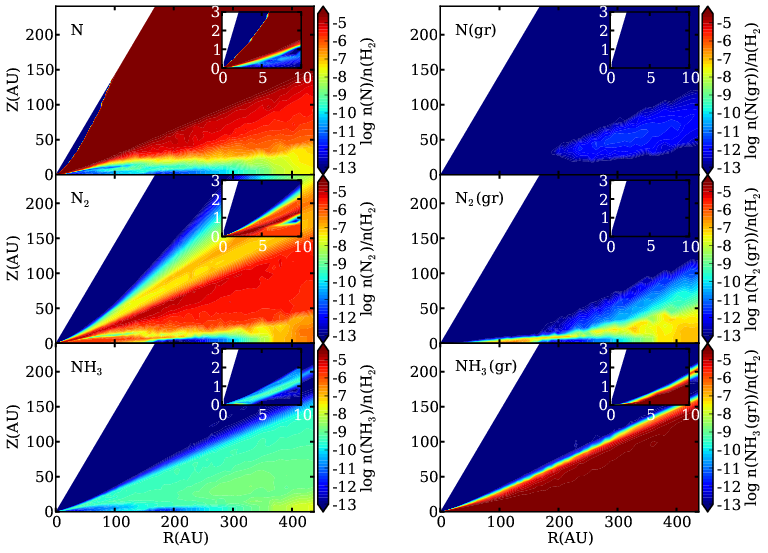}
  \caption{ Abundances relative to molecular hydrogen for Model \ammonia(gr) (part 1). The inset shows the inner disk.}
  \label{NH3grcontours1}
\end{figure}

\begin{figure}
\centering
    \includegraphics[width=1.0\textwidth]{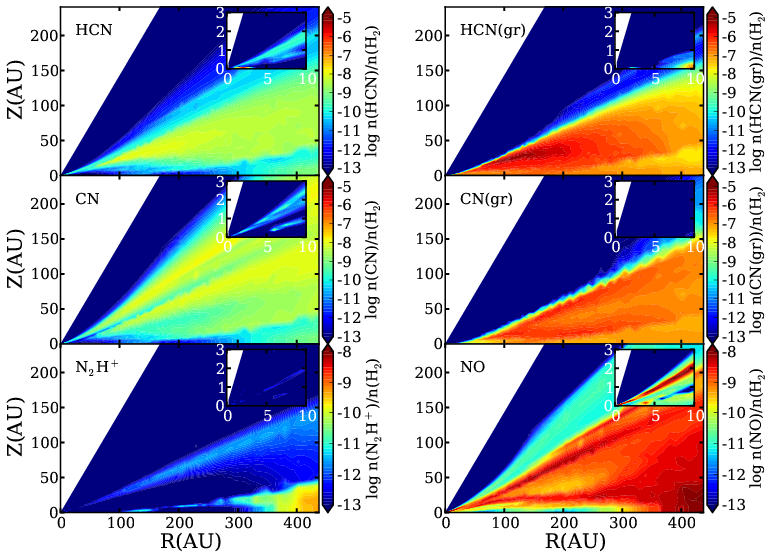}
  \caption{ Abundances relative to molecular hydrogen for Model \ammonia(gr) (part 2). The inset shows the inner disk.}
  \label{NH3grcontours2}
\end{figure}

\begin{figure}
\centering
    \includegraphics[width=1.0\textwidth]{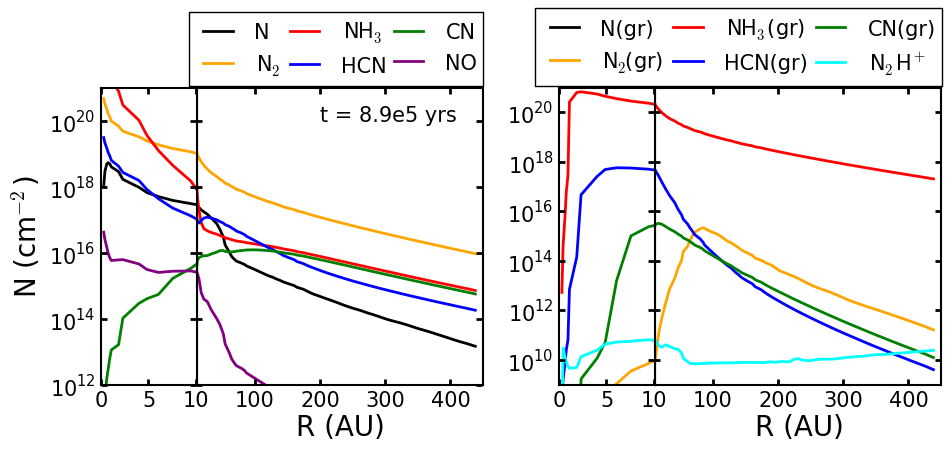}
  \caption{ Column densities for the most abundant nitrogen-bearing species in Model \ammonia(gr). The change in scaling on the x axis at 10 AU is to better show behavior in the inner disk.}
  \label{NH3grstandard}
\end{figure}

In Model \ammonia(gr) the initial gas phase atomic N has been replaced with \ammonia\ ice.
Radiation from the central star is unable to liberate the \ammonia\ except at radii less than $\sim$ 5 AU and \ammonia\ ice remains the dominant bearer of nitrogen for all times considered in our model (Figure \ref{NH3grcontours1}). Abundances of the remaining molecules are much lower than those in Model N since most of the nitrogen remains in \ammonia\ ice and does not participate in the chemistry (Figures \ref{NH3grcontours2} and \ref{NH3grstandard}).

\begin{figure}
\centering
    \includegraphics[width=1.0\textwidth]{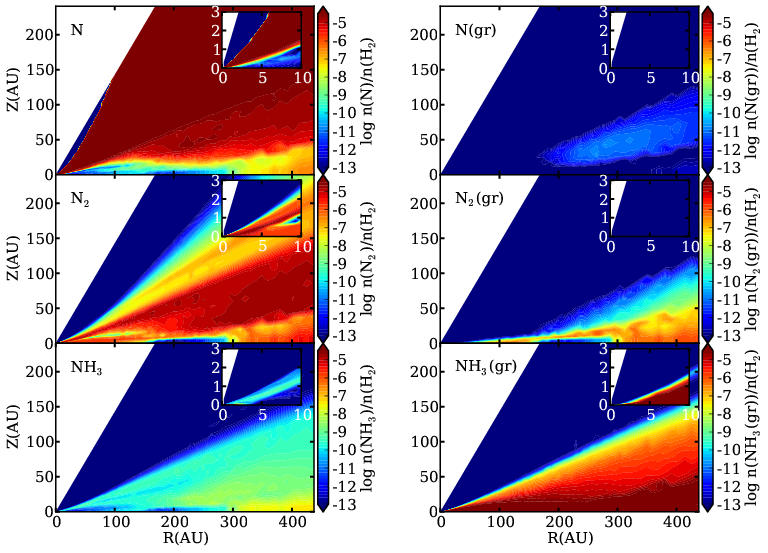}
  \caption{ Abundances relative to molecular hydrogen for Model \ammonia\ (part 1). The inset shows the inner disk.}
  \label{NH3contours1}
\end{figure}

\begin{figure}
\centering
    \includegraphics[width=1.0\textwidth]{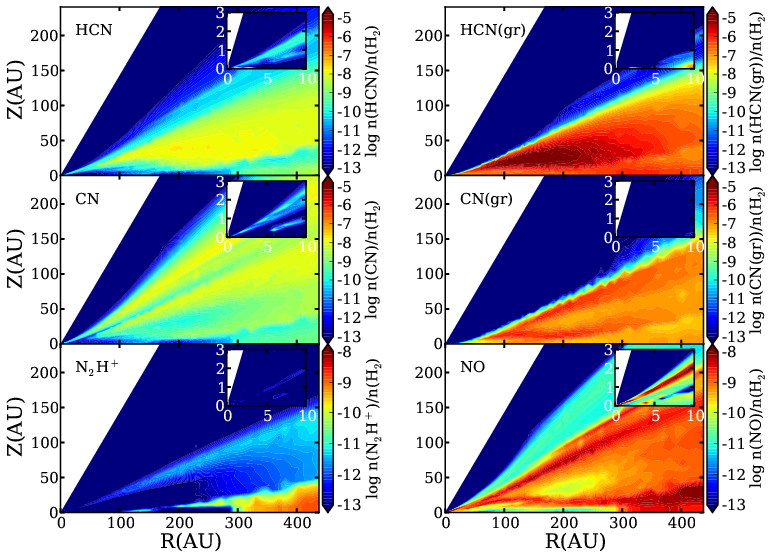}
  \caption{ Abundances relative to molecular hydrogen for Model \ammonia\ (part 2). The inset shows the inner disk.}
  \label{NH3contours2}
\end{figure}

\begin{figure}
\centering
    \includegraphics[width=1.0\textwidth]{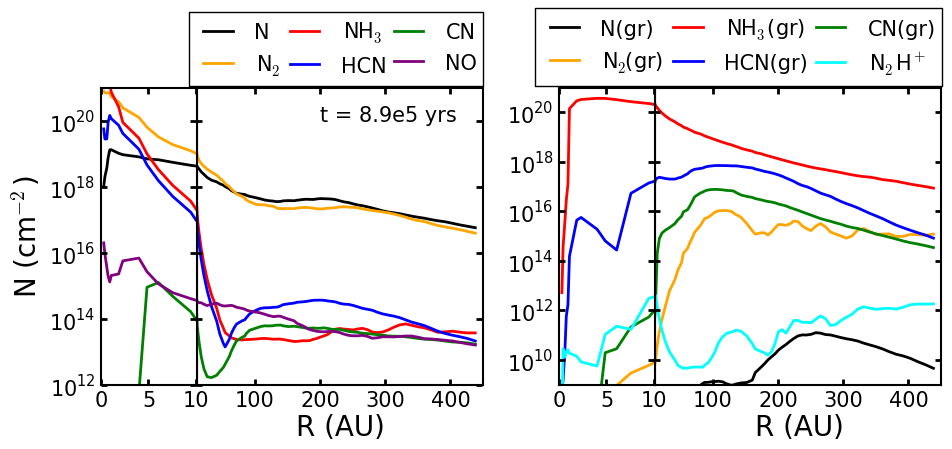}
  \caption{ Column densities for the most abundant nitrogen-bearing species in Model \ammonia. The change in scaling on the x axis at 10 AU is to better show behavior in the inner disk.}
  \label{NH3standard}
\end{figure}

Model \ammonia\ places the majority of the initial nitrogen in \ammonia\ gas. At late stages the nitrogen partitioning is very similar to that in Model \ammonia(gr), though there are several key differences (Figures \ref{NH3contours1}-\ref{NH3standard}).  \ammonia\ becomes trapped on grains in the warm molecular layer and the midplane early on. 
Before freezing out some of the gas phase \ammonia\ is able to react with other molecules. Because of this there is more \nn\ gas, as well as HCN ice and CN ice, 
in Model \ammonia\ than in Model \ammonia(gr). The formation of additional \nn\ depletes the NO, such that Model \ammonia\ has slightly lower NO abundances than those seen in Model \ammonia(gr).
In the midplane the larger \nn\ abundance also leads to the creation of more \nnhp\ beyond 300 AU when compared to Model \ammonia(gr).
In the surface layers \ammonia\ gas does not freeze out as quickly as it does closer to the midplane. Gas phase reactions are able to remove a larger fraction of the nitrogen from \ammonia, most of which becomes N and \nn\ gas.

\begin{figure}
\centering
    \includegraphics[width=1.0\textwidth]{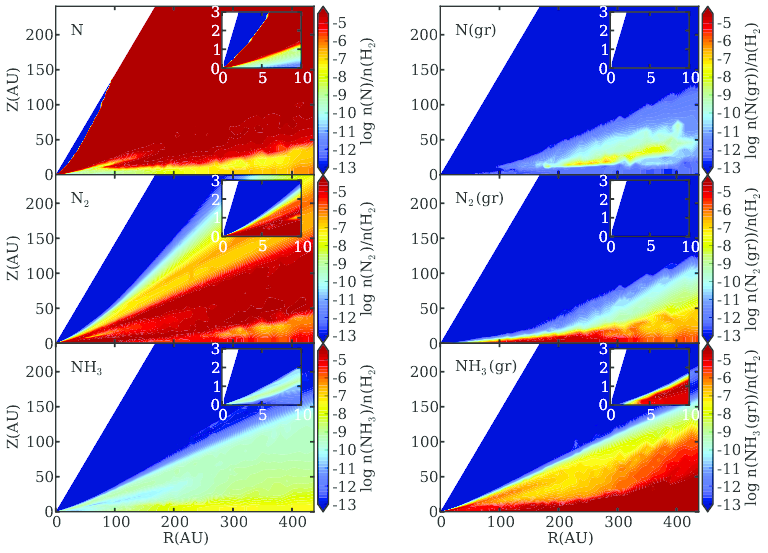}
  \caption{ Abundances relative to molecular hydrogen for Model \nn\ (part 1). The inset shows the inner disk.}
  \label{N2contours1}
\end{figure}

\begin{figure}
\centering
    \includegraphics[width=1.0\textwidth]{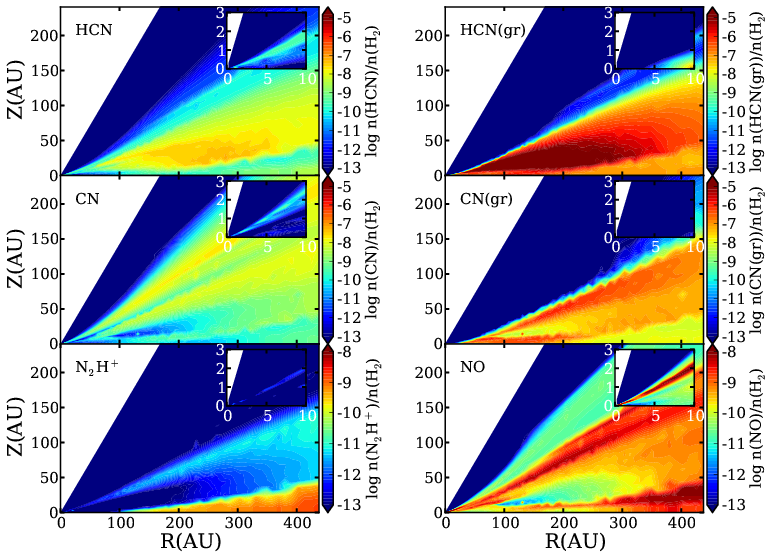}
  \caption{ Abundances relative to molecular hydrogen for Model \nn\ (part 2). The inset shows the inner disk.}
  \label{N2contours2}
\end{figure}

\begin{figure}
\centering
    \includegraphics[width=1.0\textwidth]{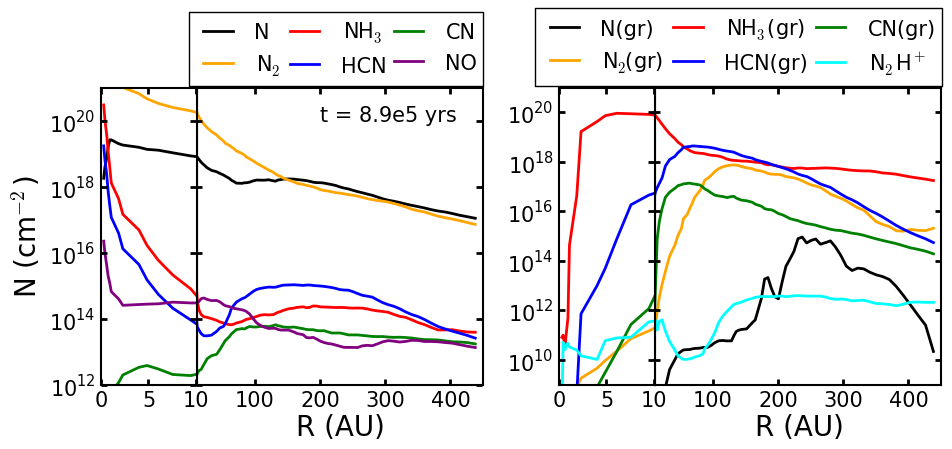}
  \caption{ Column densities for the most abundant nitrogen-bearing species in Model \nn. The change in scaling on the x axis at 10 AU is to better show behavior in the inner disk.}
  \label{N2standard}
\end{figure}

Model \nn, where most of the nitrogen is initially in gas phase \nn, results in a nitrogen partitioning very similar to that of Model N (Figures \ref{N2contours1} - \ref{N2standard}). 
That the differences are so subtle indicates that the disk is able to efficiently transfer N to \nn, assuming there is no substantial processing of N on grains limiting the availability of atomic N in the gas. Together, Model N and Model \nn\ illustrate that if the nitrogen is delivered to the disk primarily as either atomic or molecular nitrogen, the disk will likely be able to produce significant amounts of \nn\ gas via the reactions discussed previously.

\subsection{The Effects of Binding Energies}\label{highbinding}
Much of the chemistry explored in this work depends on the amount of nitrogen available for gas phase reactions. In other words, it depends on the desorption rates, which are extremely sensitive to the binding energies used. The residual midplane abundance of \ammonia\ in Figure \ref{standardcontours1} is present because the evaporation temperature of \nn\ is low enough to allow gas phase \nn\ to exist in the midplane. 
In the absence of chemical processing the gas to ice ratio is set by balancing freeze out with thermal and cosmic ray desorption:
\begin{equation}
\frac{n_{\nn}}{n_{\nn(\mathrm{gr})}} = \frac{\nu_{1} \mathrm{e}^{-E_B/T} + k_{CR}}{n_{\mathrm{gr}} \sigma v S},
\end{equation}
where $n_{\mathrm{X}}$ is the number density of species X,
$\nu_{1}$ is the vibrational frequency of \nn\ bound to the grain,
$E_B = 790$ K is the binding energy of \nn\ \citep{Oberg05},
$T = 16$ K is the dust temperature at $R=100$ AU in the midplane,
$k_{CR} = 4.37\ee{-12}$ s$^{-1}$ is the cosmic ray desorption rate,
$\sigma$ is the collisional cross section of a 0.1 micron dust grain,
$v$ is the sound speed,
and $S$ is the sticking coefficient, assumed to be 1.
 At a radius of 100 AU in the midplane this gives $n\mathrm{_{N_2}}/n\mathrm{_{N_{2}(gr)}} = 0.18$. 
 The  actual ratio for Model N is $n\mathrm{_{N_2}}/n\mathrm{_{N_{2}(gr)}} =  0.86$, indicating that chemical processing and gas phase formation have a non-negligible effect on the ratio.

\begin{figure}
\centering
    \includegraphics[width=1.0\textwidth]{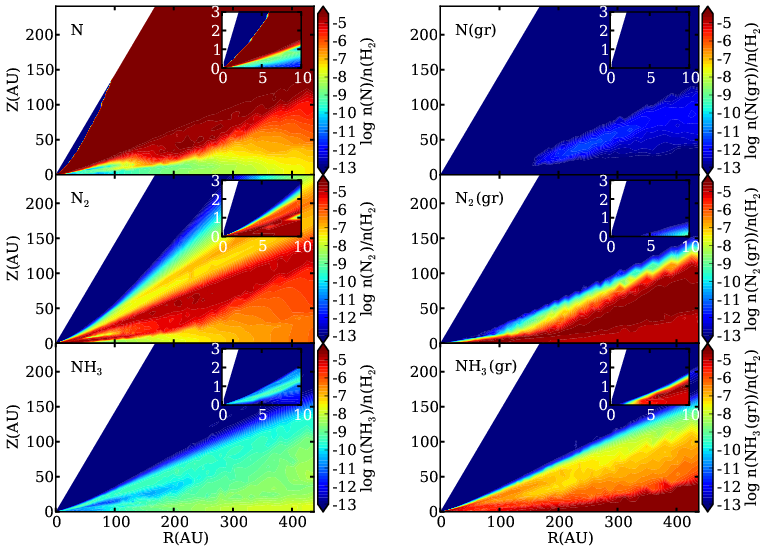}
  \caption{ Abundances relative to molecular hydrogen for Model N with and higher binding energies for CO and \nn\ (part 1). The inset shows the inner disk.}
  \label{HBcontours1}
\end{figure}

\begin{figure}
\centering
    \includegraphics[width=1.0\textwidth]{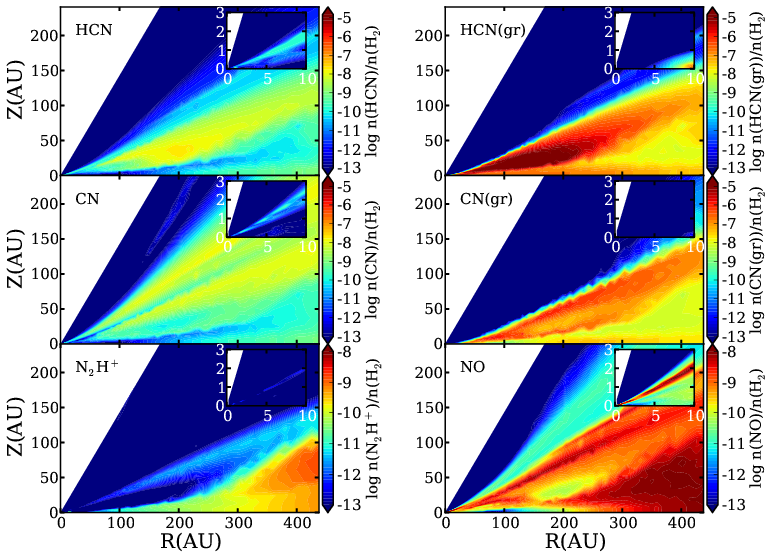}
  \caption{ Abundances relative to molecular hydrogen for Model N with higher binding energies for CO and \nn\ (part 2). The inset shows the inner disk.}
  \label{HBcontours2}
\end{figure}

Unfortunately it is not clear what assumptions should be made when determining binding energies experimentally.
The binding energy depends both on the species being bound to the grain and the composition of the grain's surface. Commonly used surfaces include CO, \water, silicates  \citep{Bergin95}, and more recently CO$_2$ \citep{Cleeves14}. 
The binding energies of molecules on ices can be calculated or measured in the lab. In Model N the binding energy of \nn\ is 790 K (corresponding to an evaporation temperature of $T_{evap} = 14.6$ K in the midplane at 100 AU) and the binding energy of CO is 855 K ($T_{evap} =  16.0$ K), as determined by \citet{Oberg05}. These binding energies assume an \nn\ and CO coated grain surface respectively and are appropriate for dust temperatures near 17 K. Figures \ref{HBcontours1} and \ref{HBcontours2} show the resulting chemical abundances when the binding energies for CO and \nn\ are both changed to 1110 K ($T_{evap} =  20.7$ K), as is appropriate for CO$_2$ coated grains and dust  temperatures between 25 K and 50 K \citep{Cleeves14}.

Changing just these two binding energies has a noticeable effect on the chemical abundances (Figures \ref{HBcontours1} and \ref{HBcontours2}).
The residual \nn\ gas near the midplane is gone, in addition to the lower abundances of N and \nn\ gas in the warm molecular layer. 
The abundances of most nitrogen-bearing ices has decreased.
However, much of the nitrogen in \nn\ gas in Figure \ref{standardcontours1} is now trapped in \nn\ ice, particularly in the outer disk. With less gas phase \nn\ available, the amount of \ammonia\ gas inside of 200 AU has also dropped. 

The total amount of \nnhp\ in the outer disk does not change, though it is less concentrated in the midplane and more abundant in the warm molecular layer compared to Model N. The outer disk contains more NO in the high binding energy model. 
With less N in the gas phase to react with, NO is not destroyed as quickly. The differences in \nnhp\ and NO are particularly interesting, as these molecules are potentially observable. 

\subsection{Tracers}
Many of the dominant nitrogen-bearing species, such as N, \nn, and all ices, are not directly observable. Instead indirect methods are needed to infer their presence. As discussed previously the existence of NO and \nnhp\ in the midplane depends on the amount of \nn\ gas present. In this section we discuss the feasibility of using NO and \nnhp\ to determine the dominant nitrogen reservoir. 

We assume a disk at a distance of 140 pc inclined at an angle of 6$^{\circ}$. Emission from the \nnhp\ J=4-3 transition, 372.67251 GHz, and NO ($4_{-143}-3_{134}$) transition, 350.68949 GHz, are calculated using LIME, a non-LTE line radiation transfer code \citep{Brinch10}. 
The \nnhp\ transition was chosen based on the \nnhp\ J=4-3 observations made by \citet{Qi13b}. The NO transition is the most readily observable transition based on RADEX calculations \citep{vanderTak07}

\subsubsection{Probes of the Distribution of Elemental Nitrogen}
\begin{figure}
\centering
    \includegraphics[width=1.0\textwidth]{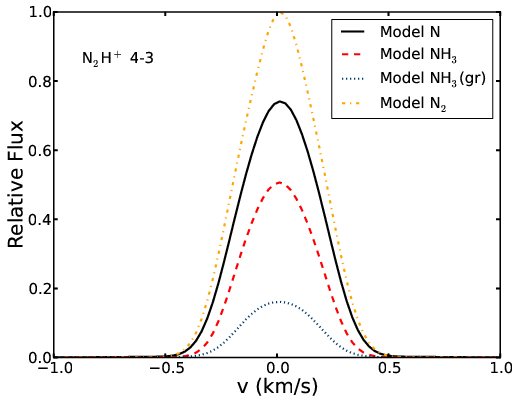}
  \caption{ Model \nnhp\ J = 3-2 line emission from a disk 140 pc away with an inclination angle of 6 degrees relative to the maximum value of 10.5 Jy.}
  \label{nnhp32}
\end{figure}

\begin{figure}
\centering
    \includegraphics[width=1.0\textwidth]{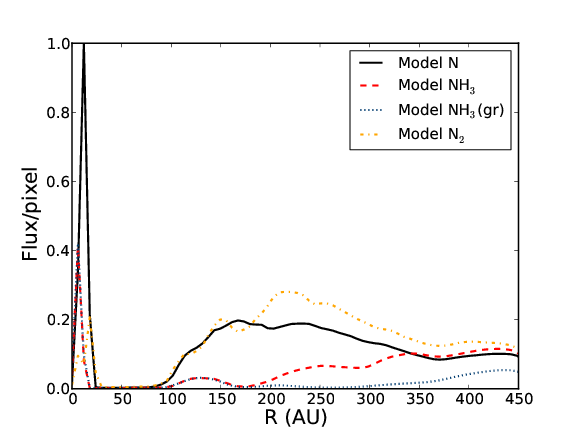}
  \caption{ Model \nnhp\ J = 3-2 line emission as a function of radius relative to the maximum value of 4\ee{-24} Jy/pixel for a disk 140 pc away.}
  \label{nnhprad}
\end{figure}

Figure \ref{nnhp32} shows the strength of the \nnhp\ (4-3) line for our four models relative to the strongest line, while Figure \ref{nnhprad} shows the average emission as a function of radius. We choose to focus on the relative line strengths because the absolute line strength is highly dependent on the physical model used. 
The strongest emission is seen in Model N in the inner 25 AU of the disk due to a temporary increase in the local \nnhp\ abundance. This is a time dependent effect that is also seen in the other models at slightly different times. As such it should not be used as a way to differentiate between the models. The cause is time dependent destruction of reactive molecules with \nnhp. These molecules, such as CO and CH$_{4}$, are depleted in local layers via ionization effects \citep{Bergin14}.

Beyond $R$ = 25 AU the strongest emission is in Model \nn.
At early times there is less atomic nitrogen available to form \ammonia\ on grains in Model \nn\ compared to Model N and a larger fraction of the hydrogen goes into the hydrogenation of carbon on grains. With more carbon in CH$_2$ there is less CO available to destroy \nnhp, leading to the stronger \nnhp\ emission in Model \nn\ beyond $R=170$ AU.
Between  $R$ = 40 AU and $R$ = 170 AU the emission in Model N and Model \nn\ is comparable. 
The remaining models, Model \ammonia\ and Model \ammonia(gr), show weaker \nnhp\ emission overall. While the emission in Model \ammonia\ and Model \ammonia(gr) is identical inside of $R$ = 150 AU, in the outer disk the amount of \nnhp\ in Model \ammonia\ increases due to the increased presence of \nn\ while the abundance in Model \ammonia(gr) remains low. This leads to overall weaker emission in Model \ammonia(gr).
The strongest \nnhp\ J=4-3 emission in Models N and \nn\ originates between $R$ = 100 AU and $R$ = 300 AU in the midplane (Figure \ref{nnhprad}), resulting in a ring of strong \nnhp\ emission, similar to that seen in TW Hya \citep{Qi13b}. 
In Models \ammonia\ and \ammonia(gr) a similar, though weaker, ring structure is seen beyond $R$ = 300 AU.

\begin{figure}
\centering
    \includegraphics[width=1.0\textwidth]{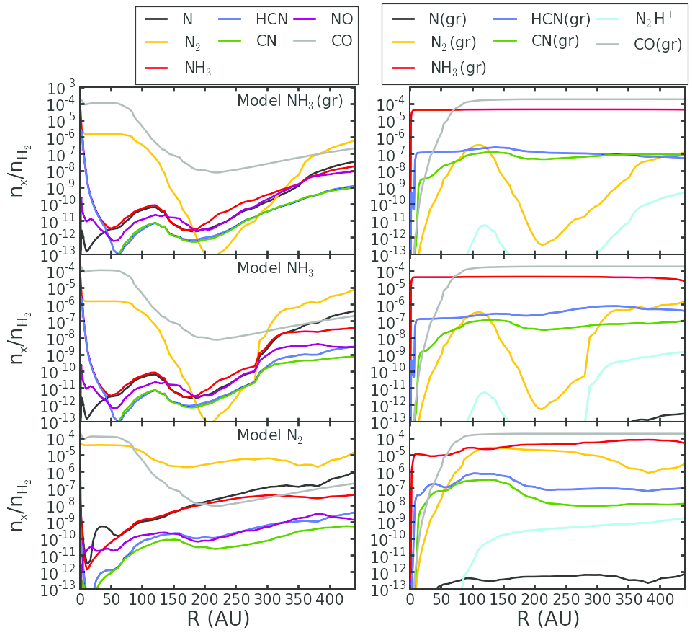}
  \caption{ Midplane abundances in Models \ammonia(gr), \ammonia, and \nn.}
  \label{snowother}
\end{figure}

Figure \ref{snowother}  illustrates the midplane abundances in Models \ammonia, \ammonia(gr), and \nn. Together with Figure \ref{snow} they illustrate the gas phase \nn\ abundance in the \nnhp\ emission region. 
 In Model N and Model \nn, the models with the strongest emission lines, at $R$ = 150 AU in the midplane, $\mathrm{ n_{N_{2}H^{+}}/n_{H_{2}}\sim 10^{-10}}$ and $\mathrm{n_{N_{2}}/n_{H_{2}}\sim 10^{-6}}$. In comparison, for Model \ammonia\ and Model \ammonia(gr),  $\mathrm{n_{N_{2}H^{+}}/n_{H_{2}}\sim 10^{-13}}$ and $\mathrm{n_{N_{2}}/n_{H_{2}}\sim 10^{-11}}$. 
 When the gas phase \nn\ abundance drops below $\mathrm{n_{N_{2}}/n_{H_{2}}\sim 10^{-6}}$, \nnhp\ is no longer present. 
Thus, the strength of the \nnhp\ line can be used as a proxy for the midplane \nn\ abundance between 100 and 300 AU for an assumed ionization rate and disk structure. This is entirely consistent with earlier work attempting to determine the \nn\ abundance from \nnhp\ in dense cores \citep{Womack92}.
\nnhp\ has been observed in several disks \citep{Dutrey07,Oberg10,Oberg11b}. Our models suggest that for \nnhp\ to be strongly emissive there must be a significant gas phase \nn\ abundance. It is likely that the mere detection of the \nnhp\ (4-3) line at the level $\sim$ Jy at 54 pc, e.g. \citet{Qi13b}, implies $n_{\nn}/n_{\hh}>\eten{-6}$. 

The distribution of NO shows some variance between our models. For all of our models it is present in the surface layers. However, in Model \ammonia(gr) and Model \ammonia\ it is also abundant in the warm molecular layer. Model N, Model \ammonia(gr) and Model \ammonia\ all show a high NO abundance near the midplane beyond $R$ = 300 AU. NO emission in strongest in Model \ammonia(gr), which has the lowest volatile nitrogen abundances. Thus, detecting NO could indicate a depleted volatile nitrogen reservoir, especially if the \nnhp\ emission from the source is weak.

\begin{figure}
\centering
    \includegraphics[width=1.0\textwidth]{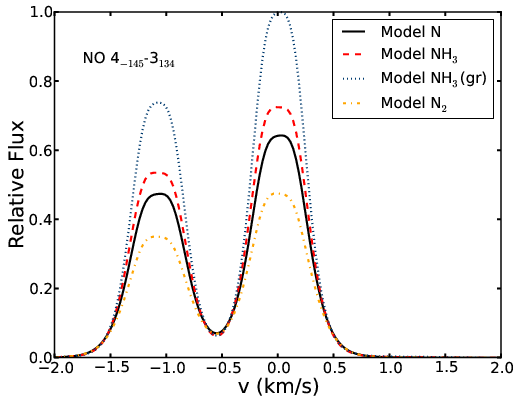}
  \caption{ Model NO  line emission from a disk 140 pc away with an inclination angle of 6 degrees relative to the maximum value of 0.43 Jy. The nearby $4_{-145} -3_{134}$ transition is also shown.}
  \label{no}
\end{figure}

Unfortunately, because NO has an uneven number of electrons ($^{2}\Pi$ ground electronic state), strong rotational transitions are replaced by a multitude of weaker lines split by $\Lambda$ doubling and hyperfine structure that is due to the non-zero spin of the nitrogen atom  \citep{Gerin92}. As a result, detecting NO is extremely difficult. Currently there are no NO detections towards protoplanetary disks, though circumstellar NO has been detected toward the embedded protostar NGC 1333-IRAS 4A \citep{Yildiz13} and it has been detected in the dense ISM \citep{McGonagle90}.
However, our models predict the NO ($4_{-145}-3_{134}$) transition (350.68949 GHz) to be on the order of several hundred mJy for a disk 140 pc away
(Figure \ref{no}). 
The strongest emission is in Model \ammonia(gr), with a flux of 437 mJy.
For the full ALMA array 4.6 minutes of integration time would be required to detect the strongest of our simulated NO ($4_{-145}-3_{134}$)  lines with a 1.4\as\ beam at the 10 sigma level assuming a spectral resolution of 0.3 km/s. We again note that our models are descriptive of general effects rather than predictive of the overall flux; however, this flux level suggests that NO may be detectable in some systems.

\subsubsection{\nnhp\ as a Probe of the CO Snow Line}
\nnhp\ is used as a tracer of the CO snow line, since gas phase CO destroys \nnhp  \citep{Qi13b}. Our models support this interpretation (Figures \ref{snow} and \ref{snowother}). The decrease in the \nnhp\ abundance between $R=200$ AU and $R=100$ AU in Models N and \nn\ corresponds with an increase in the CO gas phase abundance. In Models NH3 and  NH3(gr) the midplane \nnhp\ abundance is much lower. While there is still an increase in the midplane abundance between $R=160$ AU and $R=100$ AU in these models, the mere presence of strong \nnhp\ emission at the CO snow line suggests that \nn\ is the main nitrogen reservoir. 

\subsubsection{Surface Tracers: CN and HCN}
\begin{figure}
\centering
    \includegraphics[width=1.0\textwidth]{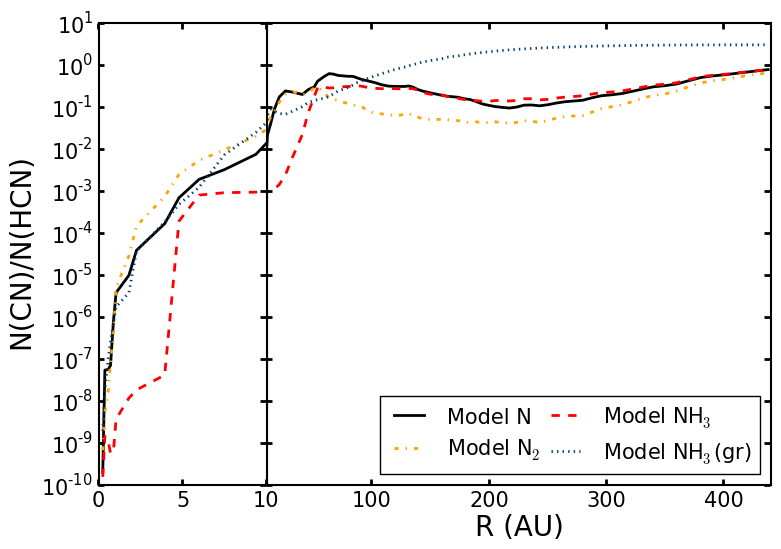}
  \caption{ Column density ratios of CN to HCN for our four models.}
  \label{CNHCN}
\end{figure}

CN and HCN are widely detected in protoplanetary disks \citep[e.g.][]{Chapillon12}. Figure \ref{CNHCN} shows the column density ratio of CN to HCN for our four models. 
In Model \ammonia\ the ratio is much larger than in the other models inside of 5 AU due to a decrease in the CN column density.
Beyond 100 AU the column density of HCN in Model \ammonia(gr) falls off, resulting in a higher CN/HCN ratio compared to the other models. 
The ratio of the simulated CN ($2_{3}-1_{2}$) and HCN (3-2) line strength is  0.5 for Model N. In comparison for most disks with detected CN and HCN emission the ratio is between 1 and 3 \citep{Oberg10, Oberg11b}. Our simulated HCN line emission is too strong. This could be due to the specific dust structure used in this model, underestimating the highly uncertain HCN binding energy, or a missing HCN processing mechanism \citep[see discussion in][]{Oberg11b}.

\subsection{Comparison to Cometary Abundances}
\begin{figure}
\centering
    \includegraphics[width=1.0\textwidth]{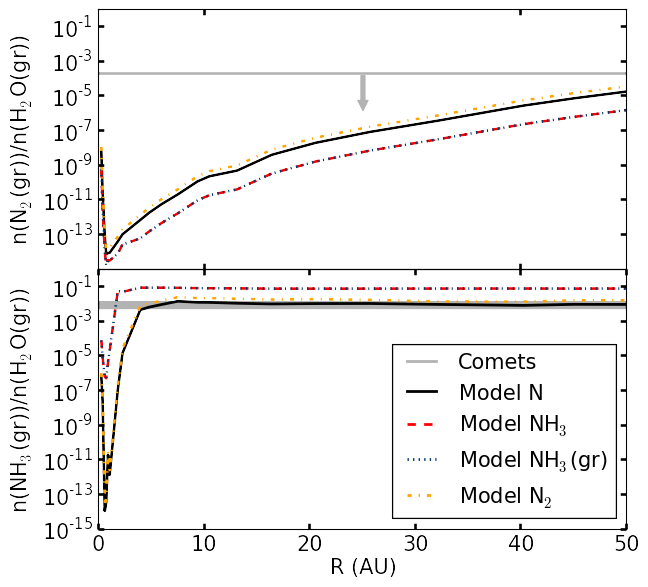}
  \caption{ Midplane abundance of \nn\ and \ammonia\ ice relative to water. The horizontal lines indicate abundances in comets.}
  \label{comets}
\end{figure}

Comets are thought to be indicative  of the abundances in the young solar nebula \citep{Bockelee11,Caselli12}.
Figure \ref{comets} shows how the \nn\ and \ammonia\ ice abundances in the midplane compare to the observed abundances in comets. The figure shows the inner 50 AU of the disk, where comets are thought to have formed in the solar system \citep{Mumma11}. 
 The cometary \nn\ value is an upper limit for Comet Halley \citep{Wyckoff91}. The \ammonia\ shows the range of values detected in Comet Hale-Bopp, Comet Halley, and Comet Hyakutake \citep{Mumma11}. The \nn\ ice abundance relative to water is below the upper limit for Comet Halley in all of our models, with Model N having the highest \nn\ to water ratio. Though our models are not meant to be analogs for the solar nebula, Model N reproduces the ammonia to water ratio observed in comets, while Model \nn\ is at the upper limit of the observed range. 
This suggests either that \ammonia\ is  processed into more refractory like material, such as part of CHON dust grains \citep[e.g.][]{Jessberger91} or \ammonia\ ice is not as easily incorporated into ices as the models predict. In this case, at face value, these models suggest N was delivered either as N or \nn.

\section{Conclusions}
We have presented four models for the initial nitrogen abundances in protoplanetary disks. Models in which the majority of the initial nitrogen is in gas phase atomic N predict that N, \nn, and \ammonia\ ice are the dominant nitrogen-bearing species at late stages, with a significant fraction of the nitrogen also in HCN ice.  When the initial nitrogen is instead placed in \nn\ gas the differences are difficult to differentiate observationally, indicating that the disk is able to convert gas phase N to \nn\ efficiently. When the nitrogen starts as \ammonia, either in the gas phase or frozen onto grains, the majority of the N remains in \ammonia\ ice. 
Model N best matches the \ammonia\ to \water\ ratio in comets, suggesting that N was delivered to the solar nebula in a highly volatile form rather than, for example, in ices. 

The presence or absence of \nnhp\ in the midplane beyond the CO snow line indicates whether nitrogen is dominated by \ammonia\ in the midplane, with the presence of \nnhp\ correlating with the presence of \nn\ gas and inversely correlating with the presence of \ammonia\ ice.
 \nnhp\ traces the snow line in all four models, though the emission is stronger for Models N and \nn. Thus the detection of strong \nnhp\ emission with a ring-like distribution suggests a disk with a high \nn\ abundance.
In addition, \nnhp\ emission can be used to determine the gas phase \nn\ abundance in the midplane.
Future sensitive observations of NO and \nnhp\ combined with disk chemical models will allow us to disentangle the nitrogen history of protoplanetary disks.

\section*{Acknowledgments}

This work was supported by funding from the National Science Foundation grant AST-1008800 and AST-1344133 (INSPIRE).

\clearpage

\clearpage

\clearpage

\clearpage

\clearpage

\clearpage

\clearpage

\clearpage

\clearpage

\clearpage

\clearpage

\clearpage

\clearpage

\clearpage

\clearpage

\clearpage

\clearpage

\begin{deluxetable}{llllll}
\tabletypesize{\scriptsize}
\tablewidth{0pt}
\tablecaption{Initial abundances relative to total H}\label{abundances}
\tablehead{
\colhead{} & \colhead{Assumed E$\mathrm{_B}$ (K){a}} & \colhead{Model N} & \colhead{Model \ammonia(gr)} & 
\colhead{Model \ammonia} & \colhead{Model \nn} \\
}
\startdata
N                      &	 800		&	2.250\ee{-5}	&	0	              &	0			&	0	\\
\nn                   & 790		&	1.000\ee{-6}      &	1.000\ee{-6} &	1.000\ee{-6}	&	1.225\ee{-5}	\\
CN                   & 1600	&	6.000\ee{-8}	&	6.000\ee{-8} &	6.000\ee{-8}	&	6.000\ee{-8}\\
HCN		      & 2050	&	2.000\ee{-8}	&	2.000\ee{-8} &	2.000\ee{-8}	&	2.000\ee{-8}\\
\ammonia      &	3080	 	&	8.000\ee{-8}	&	8.000\ee{-8} &	2.258\ee{-5}	&	0	\\
\ammonia(gr)& 	\nodata	&	0	                  &	2.250\ee{-5} &	0			&	0	\\
\enddata
\tablenotetext{a}{Binding energies are taken from the 5th release of the UMIST Database for Astrochemistry \citep{McElroy13}}
\end{deluxetable}

\end{document}